\documentclass[universe,review,accept,pdftex,moreauthors]{Definitions/mdpi} 
\usepackage{CJKutf8}
\usepackage{xspace}
\usepackage{threeparttable}

\newcommand\gaia{\textit{Gaia}\xspace}
\newcommand\fermi{\textit{Fermi}\xspace}

\newcommand\rxte{\textit{RXTE}\xspace}
\newcommand\astrosat{\textit{AstroSAT}\xspace}
\newcommand\hxmt{\textit{Insight}-HXMT\xspace}
\newcommand\integral{\textit{INTEGRAL}\xspace}
\newcommand\ixpe{\textit{IXPE}\xspace}

\newcommand\nicer{\textit{NICER}\xspace}
\newcommand\cxo{\textit{Chandra}\xspace}
\newcommand\swift{\textit{Swift}\xspace}
\newcommand\xmm{\textit{XMM-Newton}\xspace}

\newcommand{\araa}{Annu. Rev. Astron. Astrophys.}   
\newcommand{\aj}{Astron. J.}   
\newcommand{\apj}{Astrophys. J.}   
\newcommand{\apjl}{Astrophys. J. Lett.}   
\newcommand{\apjs}{Astrophys. J. Suppl. Ser.}   
\newcommand{\apss}{Astrophys. Space Sci.}   
\newcommand{\aap}{Astron. Astrophys.}   
\newcommand{\aapr}{Astron. Astrophys. Rev.}   
\newcommand{\mnras}{Mon. Not. R. Astron. Soc.}   
\newcommand{\nat}{Nature} 
\newcommand{\nar}{New Astron. Rev.}   
\newcommand{\prl}{Phys. Rev. Lett.}   
\newcommand{\pasj}{Publ. Astron. Soc. Pac.}   
\newcommand{\memsai}{Mem. Soc. Astron. Italiana}   

\newcommand\ergs{$\rm erg\,s^{-1}$\xspace}

\firstpage{1} 
\pubvolume{0}
\issuenum{0}
\articlenumber{0}
\pubyear{2024}
\copyrightyear{2024}
\externaleditor{Academic Editor: Firstname  Lastname}
\datereceived{30 October 2024} 
\daterevised{7 December 2024} 
\dateaccepted{8 December 2024} 
\datepublished{date} 
\hreflink{https://doi.org/} 

\Title{X-Ray Views of Galactic Accreting Pulsars in High-Mass X-Ray~Binaries}

\TitleCitation{X-Ray Views of Galactic Accreting Pulsars in High-Mass X-Ray Binaries}

\Author{Shan-Shan Weng $^{1,2,}$*\orcidA{} and Long Ji $^{3,}$*\orcidB{}}

\AuthorNames{Shan-Shan Weng and Long Ji}

\AuthorCitation{Weng, S.-S.; Ji, L.}

\address{%
$^{1}$ \quad School of Physics and Technology, Nanjing Normal University, Nanjing 210023, China\\
$^{2}$ \quad Institute of Physics Frontiers and Interdisciplinary Sciences, Nanjing Normal University, \newline Nanjing 210023, China\\
$^{3}$ \quad School of Physics and Astronomy, Sun Yat-sen University, Zhuhai 519082, China}

\corres{Correspondence: wengss@njnu.edu.cn (S.-S.W.); jilong@mail.sysu.edu.cn (L.J.)}

\abstract{
Accreting X-ray pulsars, located in X-ray binaries, are neutron stars with magnetic fields as strong as $B\sim10^{12\text{--}13}$\,G. This review offers a concise overview of the accretion and radiation processes of X-ray pulsars and summarizes their rich observational features, particularly focusing on complex and variable temporal phenomena, spectral properties, and evolution, the new window for X-ray polarimetry and multi-wavelength advances. We also briefly discuss other related systems, i.e., gamma-ray binaries and pulsating ultraluminous X-ray sources.
}

\keyword{X-ray pulsars; neutron stars; high-mass X-ray binaries; accretion; accretion disks; strong magnetic fields} 

\begin{document}

\vspace{12pt}

\section{Introduction}\label{intro}
As one of the brightest extra-solar X-ray sources in the sky, the X-ray binary Sco X-1 was first detected on  18 June 1962 \citep{Giacconi1962}. This groundbreaking discovery marked the beginning of X-ray astronomy. X-ray binaries (XRBs) are systems consisting of a compact object (such as a neutron star or black hole) and a companion star. Their bright X-ray emissions originate from the compact object accreting material from the donor star. XRBs are classified based on the type of compact object into black hole (BH) and neutron star (NS) XRBs \citep{Casares2017}. Alternatively, they can be grouped by the mass of the donor star ($M_{\rm co}$): low-mass X-ray binaries (LMXBs) with $M_{\rm co} < 1~M_{\odot}$,  intermediate-mass X-ray binaries (IMXBs) with $1~M_{\odot} < M_{\rm co} < 8~M_{\odot}$, and high-mass X-ray binaries (HMXBs) with $M_{\rm co} > 8~M_{\odot}$ \citep{Chaty2022}.

LMXBs are typically the old population ($\sim$$10^{9}\text{--}10^{10}$ yr), in which the compact objects accrete matter from late-type main-sequence stars, A-type stars, F-G-type subgiants and white dwarfs by Roche-lobe overflow (RLO). Most LMXBs are located in the Galactic center, bulge, and disk; a substantial fraction of LMXBs are revealed in globular clusters \citep{Grimm2002, Corral2016}. The NSs in LMXBs generally have low magnetic fields ($B \sim 10^{8}\text{--}10^{9}$ G), where the magnetosphere plays a relatively minor role in accretion. Consequently, the states of BH-LMXBs and NS-LMXBs are similar and primarily governed by the accretion rate \citep{Remillard2006, Done2007, Bahramian2023, DiSalvo2023}.

HMXBs, in contrast, represent a younger population ($\sim$$10^{5}\text{--}10^{7}$ years). Some HMXBs, such as SXP 1062 \citep{Haberl2012, Henault2012}, SXP 1323 \citep{Gvaramadze2019}, and XMMU J050722.1-684758 \citep{Maitra2021}, have been associated with supernova remnants, highlighting their status as some of the youngest XRBs, still near their birthplaces. Over 169 Galactic HMXBs are concentrated along the Galactic plane, reflecting the star formation rate and tracing the Milky Way's \mbox{structure~\citep{Liu2006, Neumann2023}}. Meanwhile, hundreds of HMXBs have been detected in nearby galaxies, especially in Small Magellanic Clouds (SMCs) and Large Magellanic Clouds (LMCs). \endnote{In this paper, we use ``Galactic accreting pulsars'' to include BeXBs in Magellanic Clouds.} Alternatively, the majority of ultraluminous X-ray sources (ULXs) are thought to be the brightest end of the HXMBs in outer galaxies \citep{Fabbiano2006, Kaaret2017,  King2023}. The spatial distribution and scaling relationships between the number and collective X-ray luminosity of HMXBs and the star formation rate further support their role as tracers of recent star formation history in their host galaxies \citep{Grimm2003, Mineo2012}.  

HMXBs are conventionally divided into supergiant X-ray binaries (SGXBs) and Be X-ray binaries (BeXBs) based on their optical companions. Generally, the former is fed directly from stellar winds (although very few of them are suggested to be disk-fed via the RLO, e.g., LMC X-4, Cen X-3, SMC X-1). BeXBs, on the other hand, are characterized by their Be star companions with circumstellar disks, which are formed by material ejected from Be stars. The compact object in these systems accretes material from both the stellar wind and the Be disk, leading to dramatic variations in their X-ray emission. 

\begin{figure}[H]
\includegraphics[width=0.6\textwidth]{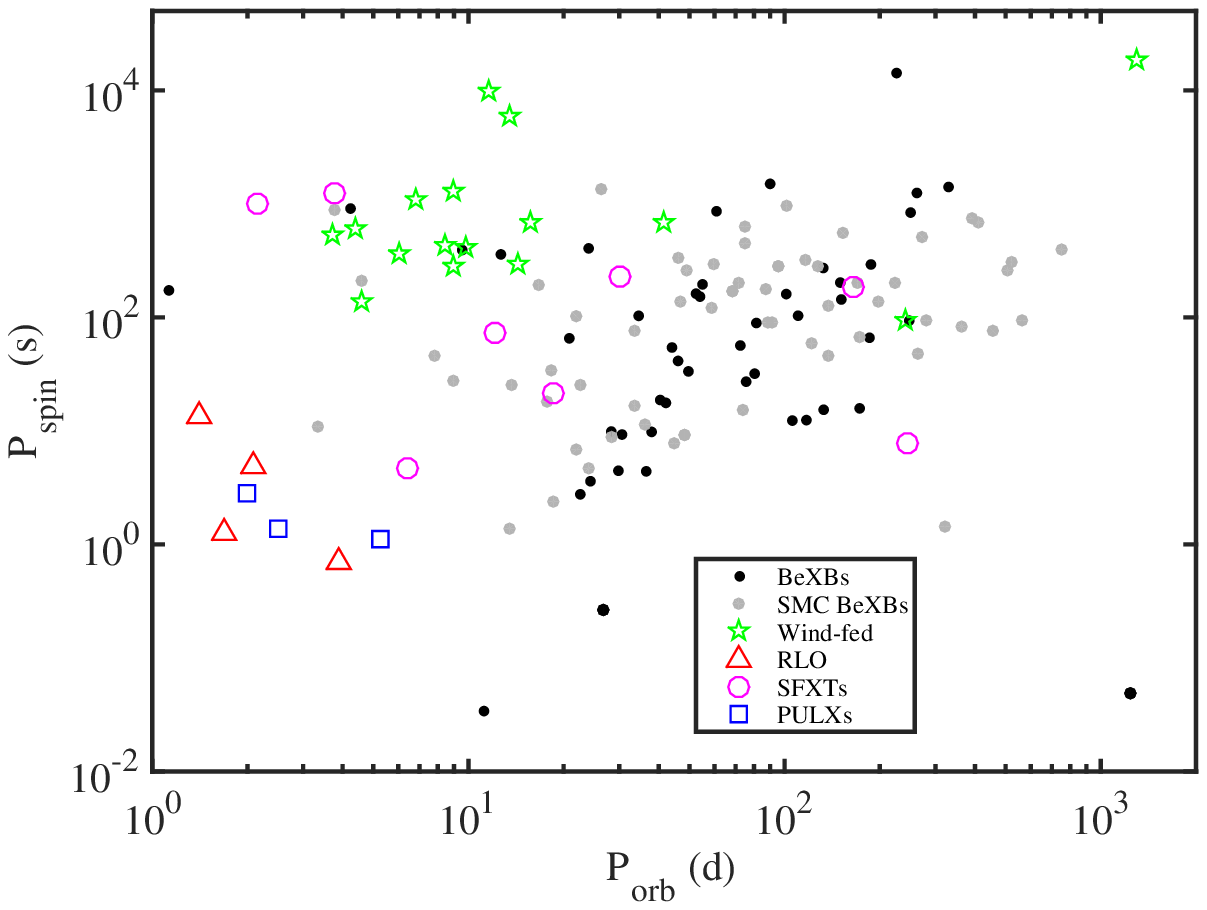}
\caption{Corbet diagram.  BeXBs in the Galaxy and Magellanic Clouds are marked with black and grey circles. The wind$-$fed SGXBs, disk$-$fed SGXBs, SFXTs, PULXs are presented with green, red, magneta, and blue symbols, respectively. Data were adopted from \citep{Liu2005, Klus2014, Weng2022,  Hou2022, Fortin2023, King2023} and the references therein. \label{fig:corbet}}
\end{figure}

To date, only about ten compact objects in HMXBs are black holes and candidates, including several wind-fed systems Cyg X-1, LMC X-1, M33 X-7, Cyg X-3, IC 10 X-1, NGC 300 X-1, HD96670 \citep{Gomez2021}, and the famous micro-quasar super-Eddington accretor SS 433.  Most other HMXBs harbor young NSs with magnetic fields of $B \sim 10^{11}\text{--}10^{13}$ G, identified through their X-ray pulsations. Different types of accreting X-ray pulsars (AXRPs, e.g., disk-fed, wind-fed SGXBs, and BeXBs) occupy distinct regions in the spin period ($P_{\rm spin}$) versus the orbital period ($P_{\rm orb}$) diagram (a.k.a. the Corbet diagram \citep{Corbet1986}, Figure \ref{fig:corbet}).

Unlike low-magnetic-field NS and BH accreting systems, the accreting processes in AXRPs are influenced not only by the accretion rate but also by the strong magnetic fields of NSs. This interplay results in distinct X-ray properties and more complex evolutionary patterns. Since the launch of X-ray telescopes in the 1970s, the accretion physics of AXRPs have been extensively studied. Several excellent reviews with different emphases on HMXBs have been published over the past decade \citep{Reig2011, Caballero2012, Walter2015, Staubert2019, Kretschmar2019, Mushtukov2022, Fornasini2023}. For the convenience of reference, we also summarize the reviews and landmark papers on various related topics in Table~\ref{tab:review}.

\begin{table}[H]
\small
\caption{Reviews and landmark papers on HMXBs.}\label{tab:review}
\setlength{\tabcolsep}{11.7mm}{\begin{tabular}{cc}
\toprule
\textbf{Description} & \textbf{Reference} \\
\midrule
Recommended reviews &  \citep{Reig2011, Caballero2012, Walter2015, Staubert2019, Kretschmar2019, Mushtukov2022, Fornasini2023, Poutanen2024} \\ 
Catalogue of HMXBs & \citep{Liu2006, Bird2010, Haberl2016, Fortin2023, Neumann2023}  \\
Achievements of {\it BATSE}, {\it INTEGRAL}, {\it Fermi}/GBM & \citep{Bildsten1997, Walter2015, Malacaria2020} \\
Formation and evolution of binaries & \citep{Grimm2003, Tauris2006, Heuvel2009, Chaty2011, Tauris2017, Vinciguerra2020} \\
Mass transfer in binary stars & \citep{Castor1975, Okazaki2001, Bozzo2008} \\
Fundamental physics in strong magnetic fields & \citep{Harding2006}\\ 
Accretion and radiation near polar caps & \citep{Basko1976, Becker2007, Becker2012, Mushtukov2015, Becker2022}  \\
Cyclotron resonant scattering features   &  \citep{Trumper1977, Staubert2019}\\
Accretion torques & \citep{Davidson1973, Ghosh1979, Davies1981, Shakura2012, Abolmasov2024}  \\
X-ray polarimetry &  \citep{Meszaros1988,Caiazzo2021,Doroshenko2023} \\
Multi-wavelength information & \citep{Reig1997, Clark1999, Coe2005, Charles2006, Chaty2008, Rahoui2008, Reig2011, Eijnden2021}  \\
\bottomrule
\end{tabular}
}
\end{table}

In this paper, we provide a general review on the X-ray observational advances and accretion physics of AXRPs, with a particular focus on the recent developments achieved by the ongoing X-ray missions, \hxmt, \nicer, \swift, \astrosat, and \ixpe \citep{Weisskopf2022}. We introduce basic physical processes in Section~\ref{model}. The temporal, spectral, and polarization characteristics of AXRPs are summarized in Sections~\ref{temporal}--\ref{polarimetry}. A brief comparison between AXRPs and related binary systems, e.g., gamma-ray binaries and pulsating ultraluminous X-ray sources (PULXs), is provided in Section~\ref{Related}. In Section~\ref{Multi}, we discuss key advances in multi-wavelength observations that provide a more comprehensive picture.

\section{Fundamental Physics in X-Ray Pulsars}\label{model}
The X-ray behavior of HMXBs is primarily determined by the characteristics of their optical companions, interactions between the accreting matter and the NS's magnetosphere as well as accretion and radiation processes occurring in the vicinity of the NS. This section provides a brief overview of the relevant core concepts and simplified physical processes underlying these systems. We encourage readers to explore more detailed and comprehensive review papers \citep{Mushtukov2022,Fornasini2023}.

\subsection{Mass Transfer in Binary Stars}
In most HMXBs, the optical donor star is either a supergiant O/B star or a Be star. For the former, the mass transfer is through accretion from the stellar wind (or the RLO in a few sources). The wind is expected to be line-driven due to the scattering of a large ensemble of spectral lines \citep{Castor1975}, and the velocity profile of the wind can be approximately described as 
\begin{equation}
v_{\rm w}(r)=v_{\infty}\left(1-\frac{R_{\star}}{r}\right)^\beta,
\label{vw}
\end{equation}
where $v_{\infty}$ is the terminal wind velocity, $R_{\star}$ is the radius of the supergiant, and $\beta \sim 0.5\text{--}1$ is the velocity gradient parameter \citep{Kudritzki1989}.
We note that the actual scenario is considerably more complex, as the wind is not perfectly smooth due to the clumpiness induced by the line-deshadowing instability \citep{Owocki1984, Owocki1988}, as well as the presence of accretion and photoionization wakes \citep{Grinberg2017}. In BeXBs, the donor star has a circumstellar decretion disk, formed by material ejected from the star's equatorial region \citep{Rivinius2013}. This disk operates as an inverse version of the Shakura--Sunyaev disk, transferring angular momentum outward.  Accretion occurs when the NS approaches or passes through the Be disk if the Be disk is large enough. This interaction may happen near periastron quasi-periodically, leading to the so-called Type-I outbursts \citep{Okazaki2001, Okazaki2013}. Conversely, the mechanism behind Type-II outbursts (Section~\ref{outburst}) remains unclear but is thought to involve highly misaligned and eccentric decretion disks~\citep{Martin2014}.

The matter can be captured by the NS only if its kinetic energy is less than the potential energy. This condition applies to inflowing matter at impact distances smaller than an accretion radius ($R_{\rm a}$), defined as 
\begin{equation}
R_{\rm a}=\frac{2GM}{v_{\rm rel}^2} \approx \frac{2GM}{v_{\rm w}^2},
\label{ra}
\end{equation}
where $M$ is the NS mass, and $v_{\rm rel} \approx v_{\rm w}$ is the relative velocity of the wind to the NS when the binary orbital velocity is $v_{\rm orb} \ll v_{\rm w}$. In addition, another crucial factor is the magnetic field ($B$) of the NS, which significantly affects the behavior of the accreted matter within the magnetospheric radius ($R_{\rm m}$). The magnetospheric radius can be estimated by balancing the magnetic pressure and the ram pressure as
\begin{equation}
R_{\rm m}=k\left(\frac{\mu^4}{2GM\dot{M}^2}\right)^{\frac{1}{7}},
\label{rm}
\end{equation}
where $\mu$ is the
magnetic dipole moment, $\dot{M}$ is the accretion rate, and $k$ is a model-dependent dimensionless parameter. Typically, $k$ is assumed to be $k \sim 0.5$ for a disk accretion and $k \sim 1$ for a spherical accretion \citep{Ghosh1979}.
{Within $R_{\rm m}$, the accreted matter is expected to be coupled with magnetic field lines and to corotate with the NS.}
If $R_{\rm m} < R_{\rm a}$, the accretion process can proceed effectively with the accretion rate given by $\dot{M}\approx \pi R_{\rm a}^2 \rho v_{\rm w}$ when $v_{\rm orb} \ll v_{\rm w}$, where $\rho$ is the wind mass density. Considering the stellar wind mass-loss rate ($\dot{M}_{\rm w}$) of the donor is
$\dot{M}_{\rm w} \approx 4\pi D^2 v_{\rm w}\rho$, where $D$ is the binary separation, the accretion rate can be described as 
\begin{equation}
\dot{M} \approx \frac{R_{\rm a}^2}{4 D^2}\dot{M}_{\rm w}.
\end{equation}
When $v_{\rm orb} \sim v_{\rm w}$, the orbital motion should be taken into account; the accretion rate is 
\begin{equation}
\dot{M} \approx \dot{M}_{\rm w} q^2 (1+q)^2 \frac{\xi \tan^4\beta}{\pi \left(1+\tan^2\beta\right)^{3/2}},
\end{equation}
where $q=M/M_{\rm donor}$ is the binary mass ratio, $\xi\sim1$, and $\tan\beta=v_{\rm orb}/v_{\rm w}$ \citep{Mushtukov2022,Davidson1973}.
On the other hand, when $R_{\rm m} > R_{\rm a}$, the accretion is significantly suppressed due to the magnetic inhibition (for details, see \cite{Bozzo2008}).

\subsection{Interactions Between the Magnetosphere and Accretion Flows}\label{magnetosphere}
If the captured matter from the wind possesses sufficient specific angular momentum, an accretion disk is expected to form outside the magnetosphere when $R_{\rm m}$ is smaller than the circularization radius ($R_{\rm circ}$), a characteristic radius within which an accretion disk exists. The circularization radius is defined as 
\begin{equation}
R_{\rm circ}=\frac{\xi^2 \omega_{\rm orb}^2 R_{\rm a}^4}{GM},
\label{circ}
\end{equation}
where $\omega_{\rm orb}$ is the orbital angular velocity of the NS, and $\xi\sim 0.2$ accounts for the angular momentum dissipation due to the density and velocity gradients in the accreting non-magnetized gas \citep{Ikhsanov2007, Ruffert1999}.
This scenario typically applies to BeXBs and may also occur in certain wind mass-loss systems. Otherwise, when $R_{\rm circ} < R_{\rm m}$, the matter  accretes spherically, which is more commonly observed in supergiant systems.

The fundamental nature of accretion is the transfer of angular momentum. Within $R_{\rm m}$, the accreted matter is forced to co-rotate with the NS due to its coupling with the magnetic field.  Efficient accretion occurs only when the NS's spin frequency  is less than the local Keplerian frequency at $R_{\rm m}$.
In other words, $R_{\rm m}$ must be smaller than the co-rotation radius ($R_{\rm co}$), which is defined as 
\begin{equation}
R_{\rm co}=\left(\frac{GMP_{\rm spin}^2}{4\pi^2}\right)^{\frac{1}{3}},
\label{rco}
\end{equation}
where $P_{\rm spin}$ is the spin period of the NS (for details about both aligned and oblique dipole cases, see \citet{Lyutikov2023}).
On the other hand, when $R_{\rm co} < R_{\rm m}$, the centrifugal barrier prevents the accretion, and the system enters into the so-called supersonic ``propeller'' regime, which is typically characterized by a low luminosity and the lack of \mbox{pulsations~\citep{Campana2002,Tsygankov2016a}}.

The interaction between the magnetosphere and the accretion flow exerts torques on the NS. 
The total torque ($N$) comprises three components: the accretion torque ($N_{\rm m}$), the magnetic torque ($N_{\rm B}$) and the radiation torque ($N_{\rm rad}$).
In the disk accretion scenario, the accretion torque is given by $N_{\rm m} = \dot{M}\sqrt{GMR_{\rm m}}$, which arises from the conservation of angular momentum of the accreted material.
This term is dominant during BeXRB outbursts when the accretion rate is relatively high.
The radiation torque can be approximately expressed as $N_{\rm rad} \sim \frac{2}{3c^3}(\mu \sin{i})^2 \Omega^3$, where $i$ is the angle between the magnetic and rotational axes, and $\Omega=\frac{2\pi}{P_{\rm spin}}$ is the NS's angular frequency. 
This torque plays a significant role in the spinning down during quiescent states.
The $N_{\rm B}$ term is more intricate, depending on how the magnetic field interacts with the accreting matter.

\begin{figure}[H]
	\includegraphics[width=1.0\linewidth]{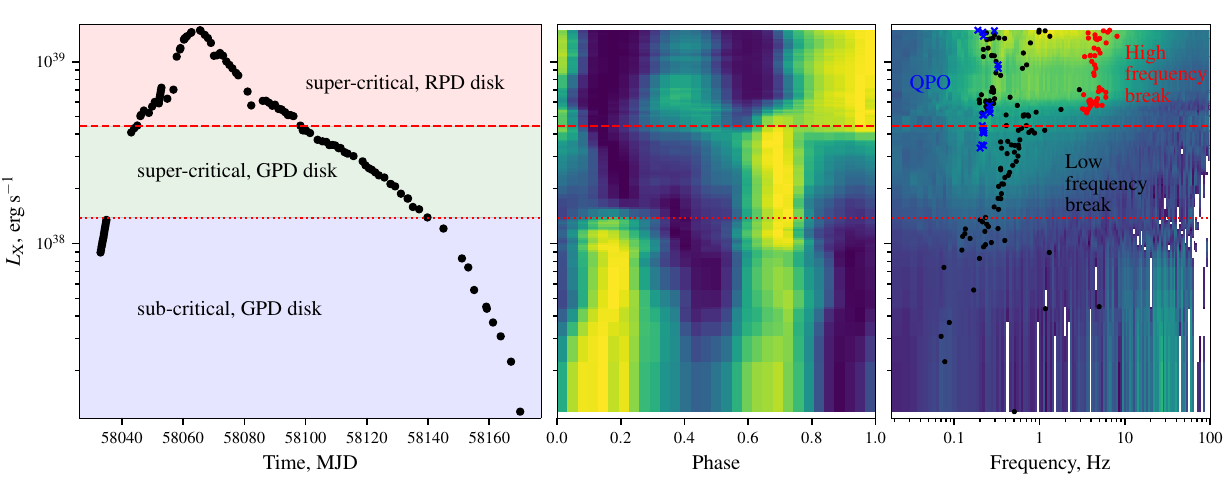}
	\caption{(\textbf{Left}) long-term lightvcurve of Swift J0243.6+6124 observed with {\it Insight}-HXMT, where color shadows present different accretion states.
 Middle: The evolution of pulse profiles, where transitions from the sub- to supercritical regimes and from the gas-pressure-dominated (GPD) to radiation-pressure-dominated (RPD) states are marked with red dashed lines.
 (\textbf{Right}) The color shows the evolution of power spectra with luminosity. The black and red points demonstrate the break and QPO frequencies measured in the power spectra.
 This figure is adopted from \citet{Doroshenko2020}.}
	\label{fig:rpd}
\end{figure}

In the disk accretion scenario, it is generally thought that the NS’s magnetic field partially penetrates into the accretion disk, generating a toroidal magnetic field and inducing torque due to the differential motion between the disk and the NS.  Assuming different prescriptions for the toroidal magnetic field, several magnetically threaded accretion models have been proposed and discussed by different authors \citep{Ghosh1979a,Ghosh1979,Wang1987,Wang1995, Rappaport2004, Kluzniak2007,Shi2015,Bozzo2018}. In addition, the interaction between the magnetosphere and the accretion disk is influenced by the disk's structure, which can result in observable phenomena in both temporal and spectral studies.  According to the accretion disk theory proposed by \citet{Shakura1973}, the disk is thin if the opacity is primarily due to electron scattering when the pressure is dominated by the gas pressure. On the other hand, the inner region of the disk might be thick at high accretion rates as the radiation pressure takes over as the dominant pressure \citep{Chashkina2017,Chashkina2019}. The disk thickness at $R_{\rm m}$ affects the size of polar caps on the NS surface, which in turn influences the observed pulse profiles (see Figure~\ref{fig:rpd}) and spectra \citep{Doroshenko2020, Ji2020, Kong2020}.   Furthermore, different accretion structures  cause different variabilities in the time domain \citep{Monkkonen2019,Doroshenko2020, LiuJR2022}.

In the case of spherical accretion in wind-fed systems, the situation is more complex and less well understood. When the accretion rate exceeds $\dot{M} \gtrsim 4\times10^{16}$\,$\rm g\,s^{-1}$, direct Bondi–Hoyle–Littleton accretion can take place, resulting in the formation of a shock above the magnetosphere. The shocked matter cools down efficiently and enters the magnetopshere via Rayleigh-Taylor instability \citep{Burnard1983,Shakura2015}. On the other hand, when \mbox{$\dot{M} \lesssim 4\times10^{16}$\,$\rm g\,s^{-1}$}, the infalling matter subsonically settles down on to the rotating magnetosphere, forming an extended quasi-static shell, and this shell mediates the angular momentum transfer to the NS. In this scenario, the accretion is less efficient and in literature it is known as the subsonic propeller regime \citep{Bozzo2008,Davies1981}. It is expected to exist within certain parameter spaces of the wind velocity, the spin and the magnetic field of the NS (for details, see Figure~2 in \citet[][]{Bozzo2008}). In this case, the radial velocity of the accreting matter is slower than the free-fall velocity and is determined by the Compton or radiation cooling rate near the magnetosphere. Within this settling accretion regime, either spinning-up or spinning-down of the NS is possible (for details, see \cite{Shakura2012}).

\subsection{Physics in the Vicinity of Magnetic Poles}\label{poles}
Within the magnetosphere, the accreted matter is guided along the dipole magnetic field lines onto the magnetic poles of the NS, where the majority of the observed X-rays are emitted.  The resulting footprint is a filled hot spot for a spherical accretion (or a hollow circle for a disk accretion), corresponding to different structures at the base of the accretion channel~\citep{Mukherjee2013, Gornostaev2021}. In theory, despite the complex coupling between matter and radiation, several approximate analytic models and simulations have been conducted by different authors \citep{Basko1975,Basko1976,Wang1981,Becker2007,Becker2012,Postnov2015,Becker2022,Zhang2022}. 

\begin{figure}[H]
	\includegraphics[width=0.9\linewidth]{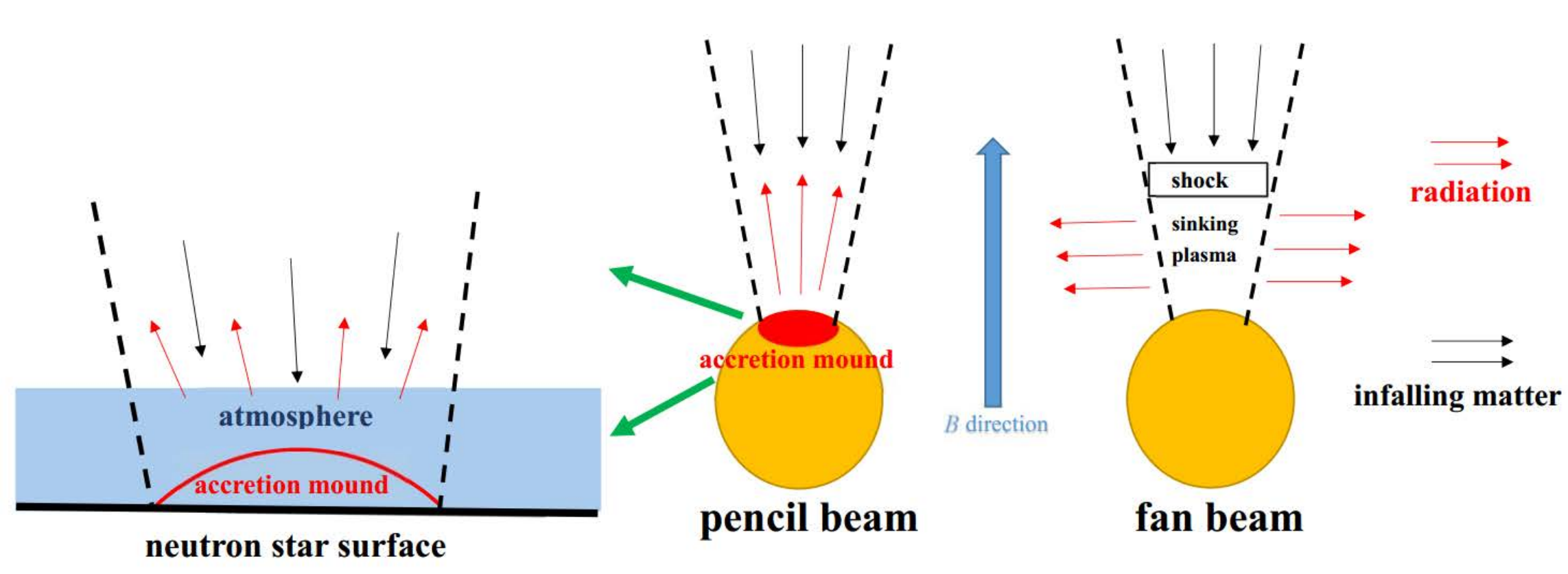}
	\caption{A sketch showing the distinct geometries and radiation patterns for the accretion mound at low accretion rates (middle) and the accretion column at high accretion rates (right). The left panel shows a zoom-in plot of the surface of the neutron star in the case of the
accretion mound. The blue arrow indicates the magnetic field direction at the pole.
The arrows on the rightmost side are the legend, i.e., black arrows present the direction of the matter movement, and red arrows show the main direction of the radiation.
}
	\label{fig:sketchbeaming}
\end{figure}

It is believed that the accretion process near magnetic poles and its observable properties mainly depend on the luminosity ($L$) and the magnetic field ($B$). When the luminosity exceeds a critical value, \endnote{$L_{\rm cirt}$ is also estimated by \citet{Mushtukov2015} considering that $L_{\rm cirt}$ is not associated with the Eddington limit and using an exact Compton scattering cross-section. The result suggests that $L_{\rm cirt}$ is not a monotonic function of $B$.} 
\begin{equation}
L>L_{\rm cirt} \sim 1.5\times 10^{37}(\frac{B}{10^{12}})^{\frac{16}{15}}\,\rm erg\,s^{-1}
\label{Lcrit}
\end{equation}
In \citep{Becker2012}, a radiation-dominated shock was found to form above the NS's surface, where the kinetic energy of the infalling matter is significantly reduced. 
Below the shock, there is a sinking zone (referred to as the accretion column), which is an extended and high-energy-density region. The column height is expected to be positively correlated with luminosity. In this case, most of the X-rays escape from the system through the ``wall'' of the accretion column, resulting in a ``fan'' beam pattern (Figure~\ref{fig:sketchbeaming}).  The radiation from the accretion column is expected to be highly polarized because the Compton scattering cross-section is energy and polarization dependent for the energy band below the cyclotron line energy, as well as the vacuum polarization effect during propagation \citep{Caiazzo2021}. However, this theoretical prediction is not fully consistent with observations (see Section~\ref{polarimetry}).
\vspace{-6pt}

When the luminosity is below $L_{\rm crit}$, the radiation pressure is insufficient to fully decelerate the infalling matter. As a result, the matter  eventually comes to rest on the NS surface via the Coulomb braking or a collisionless gas-mediated shock. It is possible that the latter is more important when $L < L_{\rm coul}\sim 1.2\times10^{37} (\frac{B}{10^{12}})^{1/3}$\,$\rm erg\,s^{-1}$ \citep{Becker2012,Langer1982}. In this regime, most of the radiation is emitted near the NS surface, where a dense accretion mound is formed, resulting in a ``pencil'' beam with the radiation emitted predominantly along the magnetic field lines. At very low accretion rates, the thermal emission from the accretion mound may serve as the primary source of seed photons for the Comptonization (see Figure~9 in \cite{Becker2022}). In addition, the overheated atmosphere, excited by the collision of infalling matter, may play a crucial role in modifying the observed spectra.  Theoretical models have been developed for luminosities in the range of $L\lesssim10^{35}$\,${\rm erg\,s^{-1}}$, considering polarization-dependent radiative transfer using different methods (for details, see \cite{Sokolova-Lapa2021,Mushtukov2021}).

\section{Temporal Properties}\label{temporal}

As accretion systems, AXRPs exhibit both regular and irregular variability across timescales ranging from seconds to years. Globally, most BeXBs are transients, whereas SGXBs tend to be persistent. In RLO SGXBs, material is steadily transferred from donor stars, forming accretion disks around the NSs, resulting in persistently high luminosities ($\sim$$10^{38}$ \ergs). In contrast, wind-fed SGXBs, which typically have narrower orbits compared to BeXBs, involve NSs embedded in dense stellar winds. Consequently, both disk-fed and most wind-fed SGXBs are persistent sources. However, a fascinating subclass of SGXBs, known as supergiant fast X-ray transients (SFXTs), has been identified, characterized by short, intense flaring emissions  (see Section \ref{temp_long}). Conversely, BeXBs have wide and highly eccentric orbit, and the radii of optical companions are smaller than their Roche lobes. NSs intermittently capture substantial amounts of material from the circumstellar disks of their companions, leading to transient outbursts (Section \ref{temp_long}). Meanwhile, a small number of BeXBs show persistent and low luminosities \mbox{($\sim$$10^{34}\text{--}10^{35}$ \ergs)}, e.g., the prototype of which is X~Per/4U~0352+309 \citep{Pfahl2002}. It is worth noting that the classification of specific sources can be dynamic, as outbursts may alter their behavior. For instance, LS~V~+44~17/RX~J0440.9+4431 was initially  categorized as a persistent BeXB \citep{Reig1999} but subsequently underwent outbursts in 2010--2011 \citep{Tsygankov2012} and 2022--2023 \citep{Doroshenko2023, Malacaria2024, Li2024}. Despite the differences in long-term behavior, various AXRP subclasses exhibit similar timing and spectral properties. In this section, we summarize their temporal evolution characteristics, including long-term trends, the defining feature of AXRPs—X-ray pulsations, aperiodic variability (e.g., quasi-periodic oscillations, QPOs), and superorbital modulation. 
 
\subsection{Long-Term Behavior}\label{temp_long}

The AXRPs in our Galaxy and in the Magellanic Clouds display a broad range of luminosities, from $\sim$$10^{32}\text{ to }10^{39}$ \ergs. They are inherently variable, showing flares and state transitions. In extreme cases, such as outbursts in BeXBs and SFXTs, their dynamic range (the ratio of peak luminosity during an outburst to quiescence luminosity) can be up to~$10^{2}\text{--}10^{7}$.

\subsubsection{Outbursts in BeXBs}\label{outburst}

Due to the occasionally rapid increase of the accretion material, BeXBs leave the quiescence state ($L_{\rm X} < 10^{35}$ \ergs) and could present two distinct types of outbursting activity, i.e., Type I and Type II outbursts. Type I outbursts are characterized by regular and (quasi-)periodic activities with a moderate intensity ($L_{\rm X} \sim 10^{36}\text{--}10^{37}$ \ergs). The peak luminosities normally arrive at or close to the binary periastron passages, and outbursts tend to cover only a small fraction of the orbit ($\sim 0.2\text{--}0.3~P_{\rm orb}$). Now, it is widely recognized that Type I outbursts are triggered as enhancement in the mass transferring from a tidally truncated Be disk around periastron passages \citep{Okazaki2001}.

Type II outbursting events (also called giant outbursts) are much rarer and occur at any orbital phase. They are known to have much brighter luminosity ($L_{\rm X} > 10^{37}$ \ergs) and last for several weeks or even for several orbital periods. Recently, some BeXBs, i.e., SMC~X-3 \citep{Weng2017}, RX~J0209.6-7427 \citep{Chandra2020, Hou2022} and Swift~J0243.6+6124 \citep{Wilson2018, Tao2019}, underwent the most violent outbursts with a peak luminosity of $\ge$$10^{39}$ \ergs, which is the threshold luminosity for the definition of ULXs. During giant outbursts, both timing and spectral properties display dramatic evolution and may go through a transition around the critical luminosity ($L_{\rm cirt} \sim 10^{37}$ \ergs, depending on the NS's magnetic field \citep{Basko1976, Becker2012, Mushtukov2015}; Equation~\eqref{Lcrit}). A quick look at the evolution of outbursts can be achieved with the hardness-intensity diagram (HID), in which the source transits from the horizontal branch to the diagonal branch and back. These two branches may reflect two different accretion modes below and beyond the critical luminosity. In the horizontal branch, AXRPs have low intensity and high X-ray variability, and the diagonal branch corresponds to high-luminosity states. The spectral and timing parameters show some significant, although not universal, correlations \citep{Reig2013}.

The operation of more-sensitive imaging X-ray telescopes (e.g., \swift, \xmm, and \cxo) allows us to monitor giant outbursts in their very low luminosity levels. Some BeXBs were observed to have the luminosity drop sharply (by a factor of more than 100) within a few days at the end of outbursts. Such a phenomenon was usually interpreted as a transition to the propeller regime, offering another critical tool for estimating the magnetic field of AXRPs \citep{Tsygankov2016, Lutovinov2017} (see Section~\ref{magnetosphere}).

\subsubsection{SFXTs}
Two decades ago, the IBIS/ISGRI instrument onboard \integral, which had a sufficiently wide field of view and high sensitivity, detected recurrent fast X-ray  transient events at low Galactic latitudes, effectively excluding the possibility of gamma-ray bursts~\citep{Sguera2005, Sguera2006}. The subsequent identification of their optical counterparts as supergiants led to the designation of this new subclass of HMXBs as  SFXTs \citep{Negueruela2006, Chaty2008, Mauerhan2010}. The hallmark feature of SFXTs is their extreme variability: they spend most of their time in a quiescence state \mbox{($L_{\rm X} \sim 10^{31}\text{--}10^{33}$ \ergs)} \citep{Sidoli2023}, and intermittently exhibit sporadic, short-lived, intense X-ray flares with durations of $\sim$$10^{2}\text{--}10^{4}$ s and peak luminosities of \mbox{$L_{\rm X} \sim 10^{36}\text{--}10^{38}$ \ergs \citep{Romano2015}}. This variability corresponds to a dramatic increase in X-ray luminosity by 2--6 orders of magnitude within a very short time span, with a low duty cycle in the bright phase (<5\%) (see \cite{Sidoli2017} for a review). Additionally, relatively fainter flares ($L_{\rm X} \sim 10^{33}\text{--}10^{35}$ \ergs) on longer timescales of days to months have also been recorded  \citep{Sidoli2008}.

SFXTs share numerous observational similarities with other known AXRPs, particularly their very hard spectral profile, which strongly suggest the presence of NSs in these systems. Indeed, both orbital and spin periods have been determined for approximately half of the known SFXTs through X-ray monitoring data \citep{Bodaghee2006, Sidoli2007, Jain2009, Drave2012, Vasilopoulos2018}. Notably, IGR~J11215-5259 displayed recurrent brief outbursts with a periodicity of $\sim$$165$ days. Each outburst lasted several days and comprised numerous X-ray flares of $\sim$$10^{3}\text{--}10^{4}$ s duration. This behavior is best explained by the presence of an elliptical orbit, with X-ray outbursts occurring near periastron, akin to the Type I outbursts observed in BeXBs but on a significantly shorter timescale \citep{Sidoli2007, Romano2009}. 

To date, around two dozen SFXTs have been discovered \citep{Fortin2023, Romano2023}. These systems likely constitute a significant fraction of HMXBs; however, their detection is hindered by several factors, including their significantly lower luminosity compared to classical SGXBs, the rarity and brevity of their outbursts, and heavy absorption \citep{Negueruela2006b, Ducci2014}. It has been proposed that SFXTs represent an evolutionary link between BeXBs and SGXBs, with more than half of SFXTs overlapping with persistent SGXBs, and some overlapping with BeXBs on the $P_{\rm spin} - P_{\rm orb}$ diagram \citep{Liu2011} (Figure \ref{fig:corbet}).

The intuitive explanation for the low luminosity in quiescence and short bright flares of SFXTs was the peculiar clumpy properties (denser and slower)  of the stellar winds and/or orbital characteristics \citep{in'tZand2005, Walter2007, Sidoli2007}. Multi-wavelength observations suggest that wind parameters play a decisive role in distinguishing classical SGXBs from SFXTs. For instance, the wind terminal velocity measured in the prototype of SFXTs, IGR~17544-2619 ($\sim$$1500$ km~s$^{-1}$) was much faster than in the classical SGXBs, Vela X-1 ($\sim$$700$ km~s$^{-1}$) \citep{Gimenez2016}. However, this interpretation remains uncertain due to a lack of comprehensive studies across a larger sample of sources \citep{Hainich2020}.  On the other hand, it has been argued that the classical Bondi-Hoyle-Littleton accretion with changes in velocity and density could only produce moderately bright flares ($L_{\rm X} \leq 10^{36}$ \ergs) as sources entered a regime dominated by strong Compton cooling \citep{Shakura2014}. Therefore, alternative mechanisms are required.

Two primary models have been proposed to explain the extreme transient behavior of SFXTs: magnetic gating mechanism and subsonic setting accretion model. In the former scenario, accretion is typically inhibited by a centrifugal barrier, with sporadic dense clumps penetrating the NS magnetosphere to trigger brief flares. Essentially, the magnetic gating mechanism is governed by the spin and magnetic field, i.e., longer spin periods require stronger magnetic fields to sustain the centrifugal barrier. More specially, magnetar-like magnetic fields ($B \sim 10^{14}$ G) have been suggested for slow rotating pulsars \citep{Bozzo2008}. Interestingly, tentative CRSFs have been reported in several SFXTs, such as IGR~J18483-0311 \citep{Sguera2010}, IGR~J17544-2619 \citep{Bhalerao2015}, and SAX J1818.6-1703 \citep{Bozzo2024}, implying relatively low magnetized NSs ($10^{11}\text{--}10^{12}$ G). However, these claims remain controversial and have not been consistently confirmed by subsequent observations  \citep{Bozzo2016}.

The subsonic accretion model was initially proposed to explain the low-luminosity AXRPs, in which a hot, quasi-static shell formed above the magnetosphere due to inefficient radiative cooling at low accretion rates (Section~\ref{magnetosphere}). The sources switch into a more effective Compton cooling regime, and matter enters magnetospheres due to the Rayleigh--Taylor instability, resulting in a moderate flare with $L_{\rm X} \leq 10^{36}$ \ergs. It was speculated that the plasma lost by the supergiant of SFXTs is non-magnetized for most of the time, while a small fraction of them carried large-scale magnetic fields transported from hot massive stars. The magnetized stellar wind triggers a much more efficient plasma entry rate via the magnetic reconnection, and the entire shell can fall on to the central NS on the free-fall time scale. That is, brighter flares can be produced by the sporadic capture of magnetized density clumps, and the magnetic reconnection triggers shell instability around the magnetosphere. Notably, this mechanism is less dependent on the NS magnetic field strength \citep{Shakura2012, Shakura2014}.

\subsection{X-Ray Pulsations}
Material from the companion is captured and channeled by the strong magnetic field onto magnetic poles of the NS surface, forming two X-ray hot spots. When there is a misalignment between the NS's rotational and magnetic axes, the observed X-ray radiation becomes pulsed as the NS spins. Therefore, investigating X-ray pulsations provides a direct access to the matter motions and emerged X-rays in the very close vicinity of pulsars.

\subsubsection{Pulse Profile}\label{pf}

\begin{figure}[H]
	\includegraphics[width=0.9\linewidth]{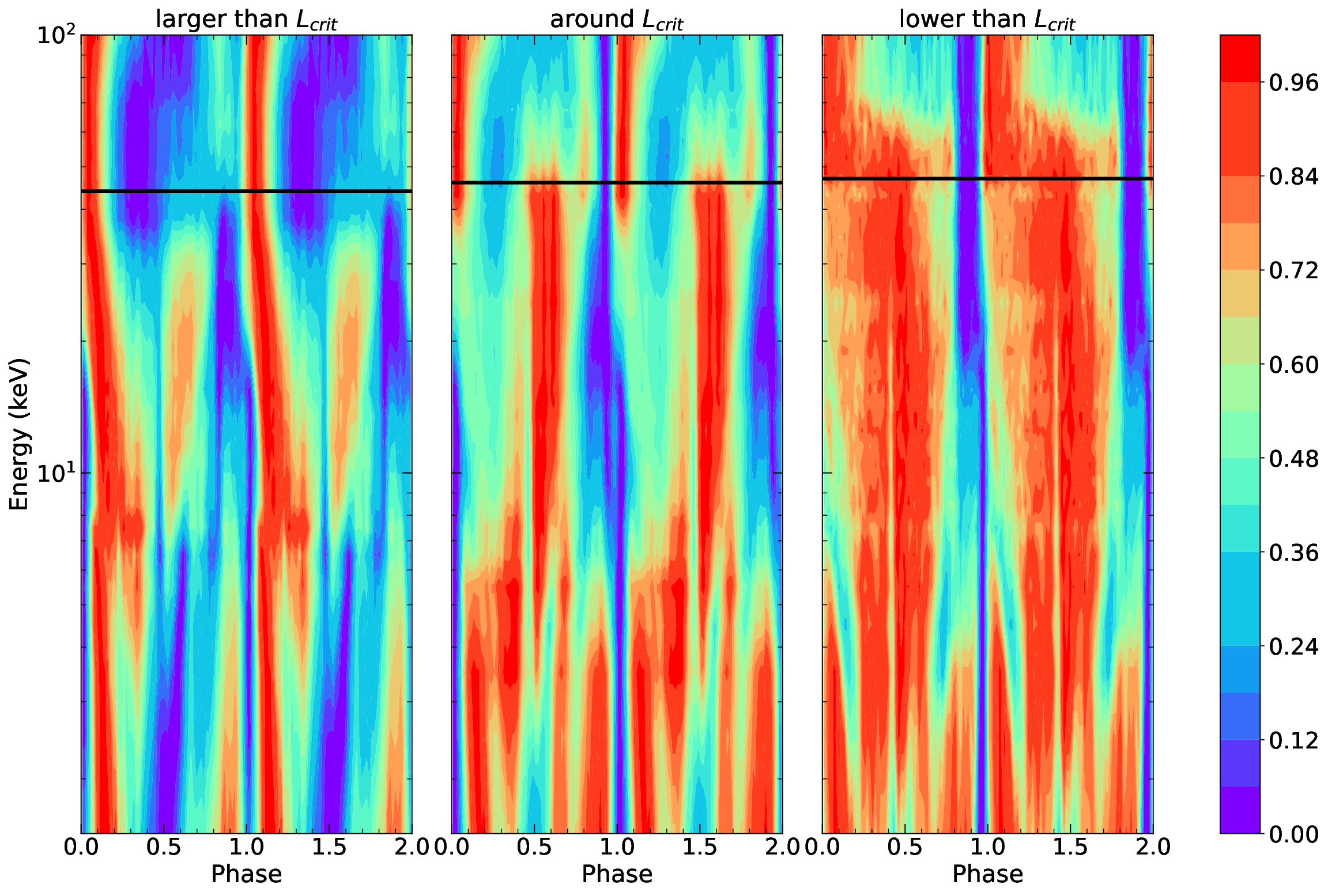}
	\caption{Pulse profiles of 1A~0535+262 plotted as a function of photon energy. From left to right, the three panels show the observations made at the peak luminosity ($L_{\rm X} > L_{\rm crit}$), around  $L_{\rm crit}$, and low luminosity ($L_{\rm X} < L_{\rm crit}$), respectively. The color bar displays the normalized intensity, and the black lines in the three panels mark the energy of the fundamental CRSFs. This plot was reproduced from \citet{Wang2022}.}
	\label{fig:a0535_profile}
\end{figure}

Ideally, the X-ray pulsations from an isotropic spot would produce a simple, symmetrical, sinusoidal-like pulse profile. However, the actual pulse shapes of AXRPs are often highly structured, vary significantly from one outburst to another, and depend on both luminosity and photon energy. For example, some sources exhibit double peaks at high luminosities and a single peak at low luminosities. This behavior is commonly interpreted as a transition between fan beam and pencil beam emission patterns \citep{Becker2012, Weng2017}. As discussed in Section~\ref{poles}, at low luminosities, radiation predominantly escapes along the magnetic axis, forming a pencil beam. In contrast, under supercritical accretion ($L_{\rm X} > L_{\rm crit}$), the accretion column becomes optically thick along the magnetic field lines. Photons then escape from the side of the column in the sinking region, producing a fan beam.  At intermediate luminosities, the pulse profile can reflect a hybrid pattern that combines pencil and fan beam components~\citep{Becker2012}. 
In reality, however, pulse profiles are far more complex. Factors such as the presence of an accretion disk, occultation and reflection by the NS surface, gravitational light bending, scattering by surrounding matter, or a distorted dipole or multipole magnetic field configuration can significantly alter the pulse shape. These modifications may result in asymmetric, narrow spikes or flare-like features. Usually, X-ray pulsations in the softer band have a more complex shape and a smaller pulsed fraction~\citep{Zhao2018}, but non-monotonic correlation is observed between the pulsed fraction and the photon energy emerged in RX~J0440.9+4431 \citep{Salganik2023}. Thanks to \hxmt, which has a very large effective area in the hard X-ray band, the pulsations were recently detected beyond 100\,keV in RX~J0209.6-7427~\citep{Hou2022}, Swift~J0243.6+6124 \cite{Zhao2024b}, and 1A~0535+262 (Figure \ref{fig:a0535_profile}) \citep{Wang2022}.

Since pulse profiles are governed primarily by the accretion process onto the NS surface, they provide valuable insights into the magnetic field geometry and X-ray beam patterns. However, the complex radiative processes, dynamics of the accretion flow, and gravitational light bending make it theoretically challenging to reproduce pulse profiles accurately \citep{Falkner2018}. Various methods and assumptions have been used to decompose pulse \mbox{profiles \citep{Parmar1989,Kraus1995,Kraus1996,Iwakiri2019,Hu2023,Saathoff2024,Thalhammer2024}}, with results that can be cross-verified using polarization \mbox{observations~\citep{Doroshenko2022, Tsygankov2022, Thalhammer2024, Zhao2024}}. On the other hand, once the beam mode switching can be determined, we will be able to estimate the critical luminosity and thus to measure the strength of the magnetic field.  For instance, the 0.25 phase offset observed in the giant outbursts of Swift~J0243.6+6124, SMC~X-3, and RX~J0209.6-7247 was interpreted as the change in photon propagation directions ($\sim$$90^{\circ}$) in the subcritical (pencil beam) and supercritical (fan beam) regimes \citep{Liu2022a}. However, we note that beam patterns at different energy bands could be different because the magnetized Comptonization cross-section is highly energy-dependent. This may lead to a large uncertainty in estimating $L_{\rm crit}$ according to the variations in the profile profiles at a narrow energy band \citep{Hou2022}.

\subsubsection{Spin Evolution}

Newborn NSs are typically fast rotators, with a spin period $P_{\rm spin} < 1$ s \citep{Popov2012, Igoshev2022}, functioning initially as rotation-powered pulsars (ejector state). In a binary, the NS is further spun down as a consequence of the interactions between the pulsar's magnetosphere and material emitted from the companion star (propeller state). When the spin period of the NS reaches a critical value, i.e., $R_{\rm co} \approx R_{\rm m}$, the centrifugal barrier cannot stop the matter from falling onto the NS surface, and the system switches on as an AXRP (Section~\ref{magnetosphere}). Accretion processes are always accompanied by an energy transformation (from gravitational energy to radiation) and matter/angular momentum transfer. AXRPs experience spin up and spin down due to the torques exerted by accreting material and may achieve an equilibrium spin rate over long times. Historically, {\it CGRO}/BATSE and \fermi/GBM have a full-sky view and have been devoted to continuously monitoring bright AXRPs in the hard X-ray window \mbox{\citep{Bildsten1997, Malacaria2020}\endnote{\url{https://gammaray.nsstc.nasa.gov/gbm/science/pulsars.html}, accessed on 9 December 2024}}. Additionally, when sources became active, \rxte, \swift, \nicer, \hxmt, \astrosat, and other satellites' pointed observations, which had a high sensitivity and a good time resolution, were triggered to trace the sources from bright to faint. Furthermore, survey campaigns aiming to discover and monitor X-ray outbursts from BeXBs in SMC were performed weekly by both \rxte \citep{Laycock2005} and \swift \citep{Kennea2018}. The gathered data have helped us to depict the decade-long behaviors of AXRPs and reveal stochastic variations and secular spin-up/spin-down evolution on long time scales. Generally, the spin-up rate ($\dot{\nu}$) and the X-ray flux are tightly correlated during the outbursts of transient sources. Specially, the $\dot{\nu}$ increases with the bolometric luminosity, following a relation of $\dot{\nu} \propto L^{b}$, with an index $b$ of 0.8--1.1 \citep{Bildsten1997, Weng2017, Liu2022a}. For persistent AXRPs, however, the relationship between accretion torque and luminosity can be more complex, including correlations, anti-correlations, or a lack of correlation altogether.

The spin evolution reflects the angular momentum exchange between accreting material and the NS magnetosphere, making it a crucial diagnostic for determining magnetic field strengths in AXRPs \citep{Klus2014, Shi2015, Weng2017, Liu2022a}. In disk accretion scenarios near the spin equilibrium (or torque reversal), magnetic fields can be estimated by equating the co-rotation radius with the magnetospheric radius. But, this assumption is not always valid as the secular spin evolution observed in many sources, in particular for the fast spin-up trend recorded in Type II outbursts. More rigorous torque models for both disk \citep{Ghosh1979, Lovelace1995, Kluzniak2007} and wind accretion \citep{Illarionov1975}, whether  close to spin equilibrium or not, have been proposed to investigate magnetic fields of AXRPs (more details in Section \ref{model}) \citep{Klus2014, Shi2015}. 

It has been well estimated that an NS can reach the longest period of $\sim$500 s by assuming a reasonable magnetic field ($B \sim 10^{12}\text{--}10^{13}$ G) and the total spin-down time in the ejector and propeller states being less than the lifetime of the young massive star \mbox{($\sim$$10^{6}\text{--}10^{7}$ yr) \citep{Urpin1998, Ikhsanov2007}}. The discovery of several relatively bright, persistent AXRPs with periods in the range of 500--10$^{4}$ s poses a serious challenge to this framework \citep{Wang2009, Haberl2012, Wang2020, Epili2024}. Various scenarios have been proposed to alleviate this contradiction, e.g., the specific quasi-spherical settling accretion model \citep{Shakura2012}, the subsonic propeller model with a normal pulsar~\citep{Ikhsanov2007, Finger2010}, and the wind-accreting magnetar model \citep{Li1999, Reig2012, Wang2020}.

\subsection{Aperiodic Variability}

Rapid aperiodic variability, driven by instabilities and fluctuation in the accretion flow, has been extensively studied in LMXBs, particularly in black hole systems. However, it remains less explored in HMXBs. Observationally, QPOs are identified and characterized by fitting power density spectra with a combination of several phenomenological Lorenztian and/or power-law functions. The quality factor $Q$ is defined as the ratio between the characteristic frequency of the component over its width ($Q = \nu/FWHM$). Broad features are referred to as noise and narrow features with $Q > 2$ are classified as QPOs. For some sources, the timing parameters of broad noise components, such as their characteristic frequency and the fractional amplitude, show significant correlations with spectral and QPOs parameters. These correlations suggest a connection between the inner accretion column and the outer accretion disk \citep{Reig2013, Li2024}.

Unlike LMXBs, which display diverse types of QPOs over a wide frequency range of $\sim$$0.1\text{--}10^{3}$ Hz across various states \citep{Wijnands1999, Remillard2006, vanderKlis2006}, QPOs in AXRPs are sporadically detected, predominantly in  transient BeXBs and occasionally in persistent sources. These QPOs typically occur at lower frequencies ($\nu_{\rm QPO} \sim 0.01\text{--}1$ Hz) \citep{Finger1996, Devasia2011, Raman2021, Liu2022b, Li2024,Manikantan2024}. During the giant outbursts of A0535+262, the center frequency of QPOs was found to be tightly correlated with the X-ray flux and spin-up rates, offering strong evidence of accretion \mbox{disk~\citep{Finger1996, Camero2012, Ma2022}}. The two most commonly used models for explaining QPOs in AXRPs are the beat-frequency model (BFM \citep{Alpar1985}) and the Keplerian frequency model (KFM \citep{vanderKlis1987}), in both of which $\nu_{\rm QPO}$ is related to the location of the inner edge of the accretion disk. But, in some cases, the QPO frequency has no obvious dependence on the X-ray flux, challenging  current \mbox{theories \citep{Raichur2008, Liu2022b}}. Interestingly, \fermi/GBM recorded the transient QPOs from RX~J0440.9+4431, which only appeared at certain spin pulse phases in the supercritical accretion regime. Such spin-phase-dependent QPOs likely originate from the bottom of the accretion column instead of the outer accretion disk \citep{Malacaria2024}.

\subsection{Superorbital Modulation}

\begin{figure}[H]
\includegraphics[width=0.7\textwidth]{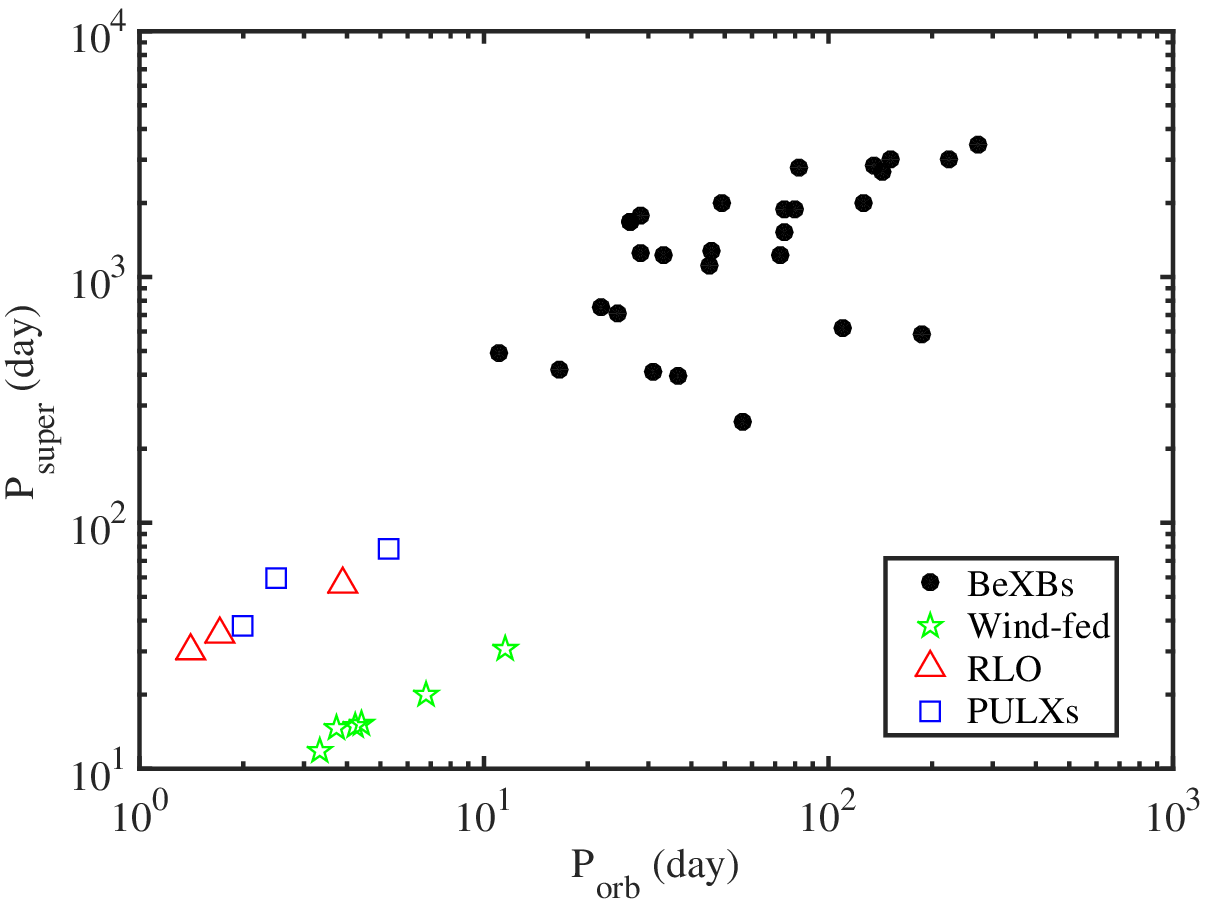}
\caption{$P_{\rm super}$ -- $P_{\rm orb}$ diagram for different types of HMXBs and PULXs. The period values of BeXBs (black circles)  were adopted from (\citet{Rajoelimanana2011, Chen2024} and the references therein), while data on wind accretion SGXBs (green pentagram) and RLO systems (red triangle) were adopted from (\citet{Corbet2013} and references therein). The data for PULXs (blue square) were adopted from (\citet{King2023} and references therein).   \label{fig:super}}
\end{figure}

Superorbital modulations on time scales of tens and hundreds of days are observed across different types of LMXBs and HMXBs \citep{Clarkson2003, Wen2006, Corbet2013}.  Diverse mechanisms have been proposed to account for the long-term variability, primarily linked to accretion processes, e.g., radiation-induced warping and/or precession disk \citep{Ogilvie2001, Pfeiffer2004, Kotze2012}, and the scenario of a precessing NS for the unique source Her~X-1 \citep{Truemper1986, Postnov2013, Doroshenko2022}. Alternatively, in BeXBs, superorbital modulation is observed at both X-ray and optical wavelengths, primarily driven by the formation and depletion of the Be star's circumstellar disk \citep{Rajoelimanana2011}. Notably, the superorbital period ($P_{\rm super}$) has a nearly linear positive correlation with the orbital period, with different types of systems occupying distinct regions in the $P_{\rm super} - P_{\rm orb}$ diagram \citep{Rajoelimanana2011, Corbet2013} (Figure \ref{fig:super}). This relationship underscores the influence of orbital dynamics on long-term variability and is further supported by the correlation between the H$\alpha$ equivalent width and $P_{\rm orb}$ \citep{Coe2005, Reig2005}.

\section{Spectral Properties}\label{spec}
The spectra of AXRPs are characterized by non-thermal radiation, resulting from bulk and thermal Comptonization scatterings of hot plasma surrounding the polar caps with seed photons emitted via
blackbody, bremsstrahlung, cyclotron processes \citep{Becker2007,Ferrigno2009,Becker2022}.
The bremsstrahlung is generally considered the primary source of seed photons in the accretion column, except for cases at very low accretion rates, where the blackbody becomes more dominant \citep{Becker2022}.
In addition, it is possible that a substantial fraction of the radiation may be emitted from a more extended region due to the reflection, which is characterized by Compton scattering in the NS atmosphere \citep{Poutanen2013, Kylafis2021}.
The spectral shape of AXRPs is highly variable, reflecting changes in the accretion structure and/or modulations caused by absorption and obscuration.
This spectral variability can be explored via spectral modeling or a model-independent hardness-intensity diagram \citep{Reig2013}.
In observations, the spectra of AXRPs generally consist of a continuum component and several narrow features, which are discussed in detail below.

\subsection{Continuum}
A non-saturated Comptonization results in a power-law-like spectral shape, which cuts off at the plasma temperature. Therefore, some phenomenological models, such as the cut-off power-law ({\it Cutoffpl}), the high-energy cut-off power-law ({\it Highecut}), the negative and positive exponential cut-off ({\it Npex}), and the Fermi Dirac cut-off ({\it Fdcut}), have been widely used in the literature to describe the continuum photon spectra of AXRPs \citep{Mihara1998,Coburn2002, Weng2019}. Their shapes are presented as follows:\vspace{6pt}
\begin{align}
Cutoffpl(E; K, \Gamma, E_{\rm fold})=K\,E^{-\Gamma} \exp(-E/E_{\rm fold})	\\
Highecut(E; K, \Gamma, E_{\rm cut}, E_{\rm fold})=K\,E^{-\Gamma} \left\{
\begin{aligned}
	1,\ if\ E \leq E_{\rm cut}\\
	\exp(\frac{-E-E_{\rm cut}}{E_{\rm fold}}),\ if\ E\geq E_{\rm cut} \\
\end{aligned}
\right.\\
Npex(E; K_1, \Gamma_1, K_2, \Gamma_2, E_{\rm fold})=(K_{1}\,E^{-\Gamma_1} + K_{2}\,E^{+\Gamma_2}) \exp(-E/E_{\rm fold})\\ 
Fdcut(E; K, \Gamma, E_{\rm cut}, E_{\rm fold}) = \frac{K\,E^{-\Gamma}}{1+\exp(\frac{E-E_{\rm cut}}{E_{\rm fold}})}
\end{align}
where $E$ is the photon energy, and $K, \Gamma,  K_1, \Gamma_1, K_2, \Gamma_2, E_{\rm cut}, E_{\rm fold}$ are free parameters. Since these models exhibit a similar shape, a given spectrum can often be adequately fitted by multiple models.
There are also several physics-motivated models that have been employed to describe the spectra of AXRPs, such as {\tt compTT}\endnote{\url{https://heasarc.gsfc.nasa.gov/xanadu/xspec/manual/XSmodelComptt.html}, {accessed on 9 December 2024}}, {\tt comptb}\endnote{\url{https://heasarc.gsfc.nasa.gov/xanadu/xspec/models/comptb.html}, {accessed on 9 December 2024}},
{\tt compmag}\endnote{\url{https://heasarc.gsfc.nasa.gov/xanadu/xspec/manual/node155.html}, {accessed on 9 December 2024}}, and {\tt bwcycl}\endnote{\url{https://heasarc.gsfc.nasa.gov/xanadu/xspec/manual/XSmodelBwcycl.html}, {accessed on 9 December 2024}}.
{\tt compTT} is an analytic thermal Comptonization model \citep{Titarchuk1994}, and {\tt comptb} is more complex, which takes both thermal and bulk Comptonizations into account \citep{Farinelli2008}.
{\tt compmag} and {\tt bwcycl} are two models developed specifically for the accretion column of highly magnetized X-ray pulsars assuming different velocity profiles of the infalling material \citep{Farinelli2012, Becker2007, Ferrigno2009}.
They can be used to constrain the physical properties of the accretion column (i.e., the radius, the optical depth, and the electron temperature) and have been successfully applied to many sources, e.g., Her X-1 \endnote{We consider Her X-1 in this paper although it is an IMXB \citep{Reynolds1997}, because it shares a lot in common with the other AXRPs in HMXBs.}
, RX J0440.9+4431, Cen X-3, and 4U 1538-522 \citep{Wolff2016, Ferrigno2013, Farinelli2016, Thalhammer2021, Hu2024}.
Regardless of the model used, the spectral parameters typically show a pronounced dependence on luminosity. 
For most strong outbursts of AXRPs, the spectral evolution with luminosity reveals two distinct branches (see Figure~6 in \cite{Reig2013}), akin to the behavior observed in the hardness-intensity diagram.
This can be interpreted in terms of the different accretion regimes discussed in Section~\ref{poles}.
We caution that luminosity is not the only factor that can impact the underlying physics.
For example, the hysteretic effect—characterized by spectral and timing differences between the rising and declining phases of outbursts at similar luminosity states—has been observed in many sources \citep{Reig2008,Doroshenko2017,Wilson2018,Wangpj2020,Kong2021}.

\begin{figure}[H]
	\includegraphics[width=0.65\linewidth, angle=-90]{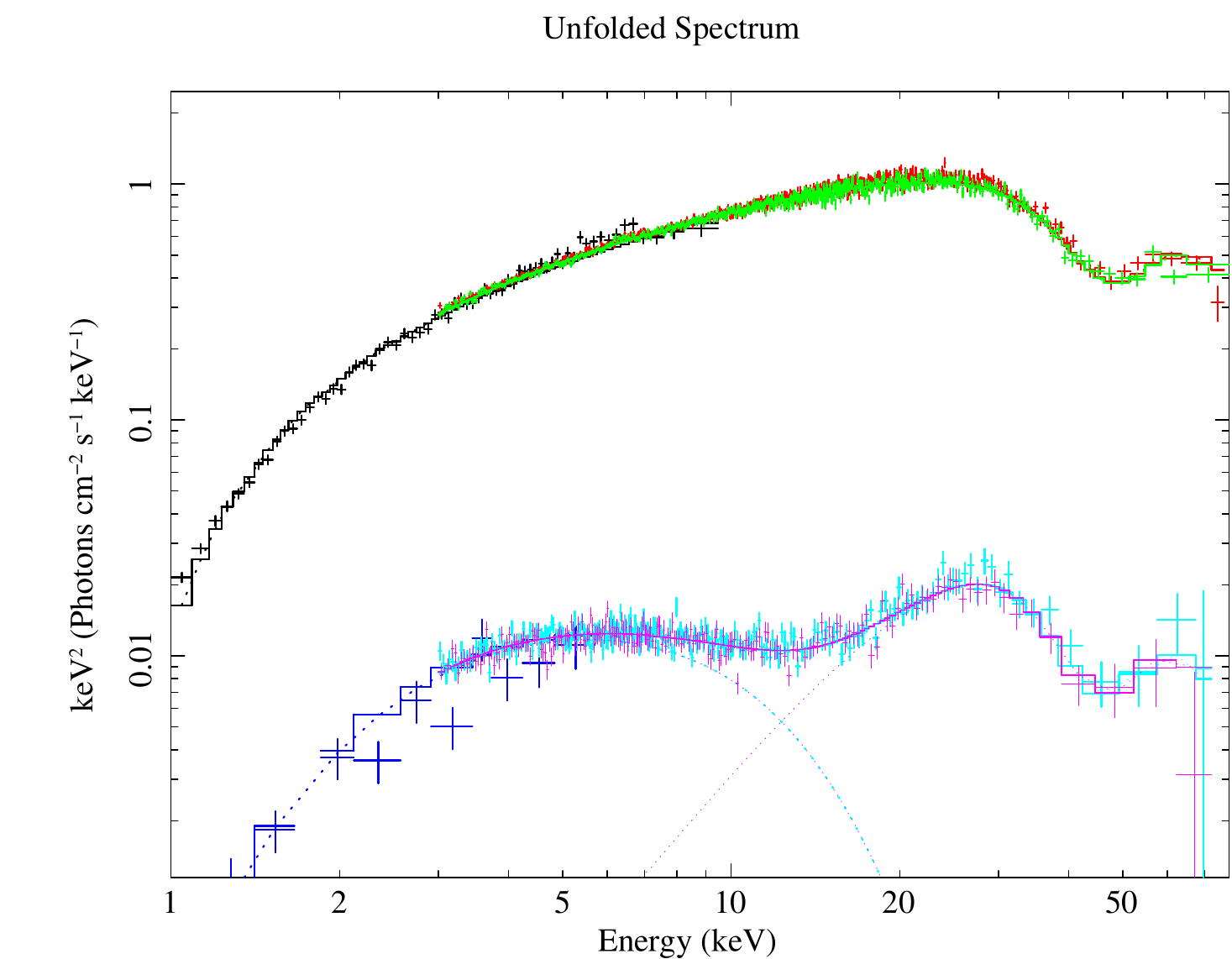}
	\caption{\textls[-15]{Representative spectra of AXRPs at high- and low-luminosity states. The structure around 45\,keV is caused by the cyclotron line. This plot was reproduced using the data from \mbox{\citet{Tsygankov2019b}}.}}
	\label{fig:spectra}
\end{figure}

The above models have limitations when fitting the spectra obtained from low-luminosity sources because they only consider the photons escaping through the wall of the accretion column, which is dominant when the accretion rate is relatively high.
For the faint states of GX 304-1, 1A 0525+262, and X Persei observed at a low luminosity of $\sim$$10^{34}\,{\rm erg\,s^{-1}}$, their spectra present a double-hump shape, instead of a cutoff power-law shape as mentioned \mbox{above \citep{Tsygankov2019a, Tsygankov2019b, Doroshenko2012}} (Figure~\ref{fig:spectra}).
This spectral evolution at low luminosities is also associated with changes in pulse profiles \citep{Xiao2024}.
Theoretical calculations suggest that the soft thermal hump at low energies originates from the deep layer of the NS atmosphere, and the high-energy hump might be explained as resonant Comptonization in the heated non-isothermal part of the atmosphere and cyclotron photons \citep{Sokolova-Lapa2021, Mushtukov2021}.
To the best of our knowledge, the only physical model currently available for data analysis is an updated version of {\tt bwcycl} ({\tt BW22}), which has been successfully applied in RX J0440.9+4431 \citep{Becker2022, Li2024}.

In addition to the Compton component, other broad features are observed in AXRPs' spectra.
For example, a soft excess commonly appears below 1\,keV, which can be modeled with a blackbody component \citep{Hickox2004}. 
It may originate from different emission processes, depending on the luminosity, such as the reprocessing of hard X-rays from the NS by the inner region of the accretion disk and emission from the surrounding diffuse gas. 
Furthermore, in many cases, the spectral fitting is not perfect, and there might be some residuals around 10\,keV (known as the ``10-keV-feature''; \citep{Coburn2002}).
They can be technically alleviated by including an additional broad emission or absorption line \citep{Manikantan2023}.
Their physical origin is still unclear.
It is also possible that they just reflect the imperfections of the available continuum models.

The radiation observed from Earth is absorbed by both the interstellar medium (ISM) and the matter surrounding the binary system. Given that the latter might be inhomogeneous and highly clumpy, the absorption appears to be highly variable in many sources. When the equivalent hydrogen column ($N_{\rm H}$) is high enough, the absorption significantly influences the soft part of the continuum as $I_{\rm obs}(E)=I_{\rm source}(E) e^{-\sigma(E)N_{\rm H}}$, where $\sigma(E)$ is the abundance-dependent photoelectric cross-section \citep{Wilms2000}. If only a fraction of the emergent photons is absorbed while the rest are detected directly by observers, a partial covering model (e.g., {\tt pcfabs\endnote{\url{https://heasarc.gsfc.nasa.gov/xanadu/xspec/manual/XSmodelPcfabs.html}, {accessed on 9 December 2024}}}) is usually employed. Studying absorption variations at different orbital phases can provide valuable information about the surrounding large-scale accretion structures, such as the accretion wake \citep[][]{Suchy2008,Islam2014, Grinberg2017,Balu2024}. In addition, with high-energy-resolution instruments, such as the gratings on {\it Chandra} and {\it XMM-Newton} and the microcalorimeter on {\it XRISM}, the velocity and structure of fast outflows, i.e., the accretion disk wind, can be resolved through blue-shifted narrow absorptions  \citep{Reynolds2010, Miller2011, Kosec2020, Kosec2023}. Furthermore, after the absorption, the re-emission of X-rays (known as ``X-ray reprocessing''), e.g., Fe $K_{\alpha}$ fluorescence, also reflects the distribution of the material, being an important diagnostic tool to probe the environment in binary systems \citep{Torrejon2010,Tzanavaris2018, Aftab2019, Ji2021}.

\subsection{Cyclotron Resonant Scattering Features}
In many AXRPs, there are broad absorption features seen in the spectra, which are known as cyclotron resonant scattering features (CRSFs) or cyclotron lines. They are caused by the resonant scatterings of photons between discrete Landau levels of electrons' motion perpendicular to the strong magnetic field lines. Since the first discovery in Her X-1 in 1976 through a balloon experiment \citep{Trumper1977,Truemper1978}, CRSFs have been detected in several tens of sources using broadband observatories, such as {\it BeppoSAX}, {\it RXTE}, {\it INTEGRAL}, {\it Suzaku}, {\it NuSTAR}, and Insight-{\it HXMT} (for a detailed review, see \citep{Staubert2019}). These features are of significant importance as they provide a model-independent way to directly measure the magnetic fields of AXRPs via the ``12-B-12'' rule: $E_{\rm cyc}\approx\frac{n}{1+z}11.6\times B_{12}$\,keV, where $E_{\rm cyc}$ is the centroid line energy, $z$ is the gravitational redshift, $B_{12}$ is the magnetic field in the line-forming region in units of $10^{12}$ G, and $n$ is the quantum number. In some sources, multiple cyclotron lines can be observed, e.g., 4U 0115+63 and GX 301-2 \citep{Heindl1999, Santangelo1999, Furst2018}, while their line energies might  not always be harmonically spaced \citep{Liu2020,Roy2024,Zalot2024}. Cyclotron lines in AXRPs are broad features due to the Doppler effect for thermal broadening \citep{Meszaros1985} and/or a gradient of the magnetic field in the line formation region. In observations, the CRSF width is typically a few keV and shows a correlation  with $E_{\rm cyc}$ \citep{Coburn2002}.
This inevitably results in the slight coupling between the CRSF model and the continuum model.

Theoretical studies of CRSFs have been conducted using both analytical methods and Monte Carlo simulations by different authors assuming different line-forming regions and configurations \citep{Nishimura2008,Nishimura2011,Nishimura2014,Nishimura2019,Nishimura2022,Schwarm2017a,Schwarm2017b,Kumar2022,Kylafis2021,Loudas2024}. Due to the complexity of physically motivated models, phenomenological models are often employed in observational studies to fit the data. The most commonly used models in the literature are multiplicative models {\tt gabs}\endnote{\url{https://heasarc.gsfc.nasa.gov/xanadu/xspec/manual/XSmodelGabs.html}, {accessed on 9 December 2024}} and {\tt cyclabs}\endnote{\url{https://heasarc.gsfc.nasa.gov/xanadu/xspec/manual/node247.html}, {accessed on 9 December 2024}}, where the former has a Gaussian profile, and the latter is pseudo-Lorentzian. Both models are characterized by the centroid energy, the line width, and the strength (or the optical depth) of the line. Due to the broad nature of CRSFs, the selection of continuum and line models introduces systematic uncertainties in $E_{\rm cyc}$, typically by a few keV.

\begin{figure}[H]
	\includegraphics[width=0.73\linewidth]{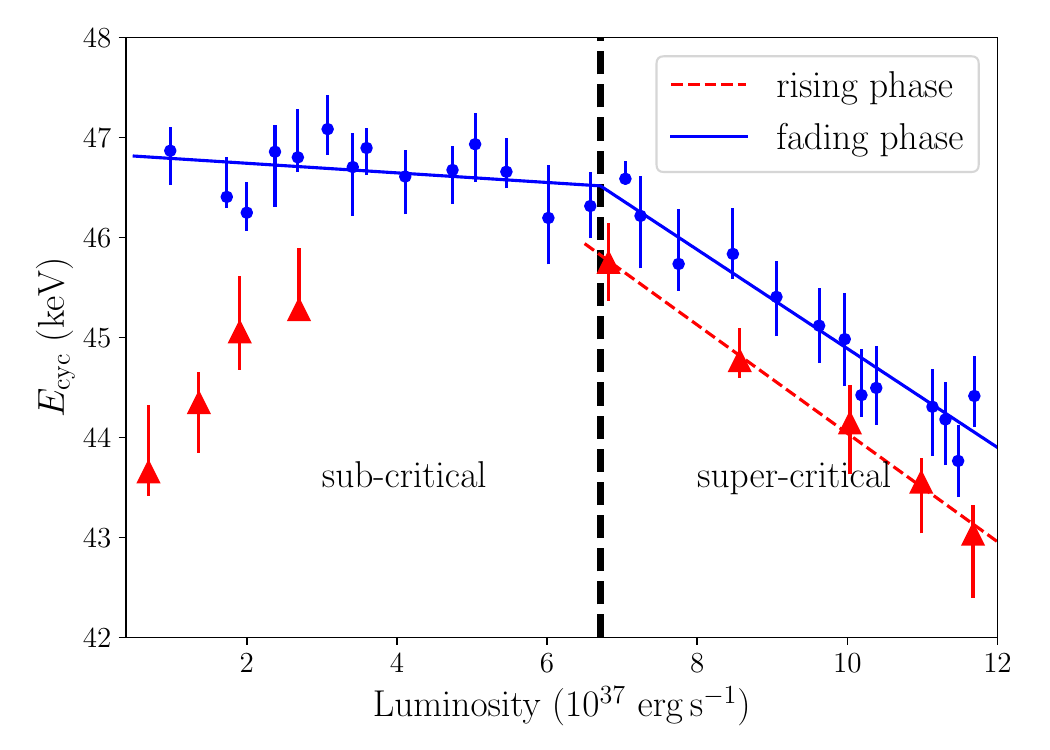}
	\caption{The evolution of $E_{\rm cyc}$ as a function of luminosity ($L$) observed in 1A 0535+262. Red triangles and blue points present the data from the rising and fading phases of the 2020 giant outburst.
	Their discrepancy reflects the hysteresis effect.
	The black dashed line represents the critical luminosity $L_{\rm crit}$, which roughly separates two accretion regimes.
	Below $L<L_{\rm crit}$, $E_{\rm cyc}$ remains stable or exhibits a positive correlation with luminosity, while above $L_{\rm crit}$ a negative correlation  when $L>L_{\rm crit}$.
    This figure was reproduced based on the data from \citet{Kong2021}.
}
	\label{fig:kong2021}
\end{figure}

It is important to note that the observed $E_{\rm cyc}$ does not directly reflect the intrinsic polar magnetic field but rather measures the local magnetic field where resonant scattering occurs. Consequently, $E_{\rm cyc}$ is expected to vary depending on the line-forming region.
In observations, $E_{\rm cyc}$ is typically not constant, showing variations with both the spin phase and luminosity.
The spin-phase-dependent line is related to the distortion of the local magnetic field of two poles due to the confinement of the accreted matter
\citep{Mukherjee2012}.
Additionally, in some cases, the CRSF is only detectable at certain spin phases and may not appear in the phase-averaged spectrum, such as the highest CRSF discovered in Swift J0243.6+6124~\citep{Kong2022}.
The relationship between $E_{\rm cyc}$ and luminosity was initially investigated in Her X-1 \citep{Staubert2007} and has been reported in other sources (for a summary, see Figure~9 in \citep{Staubert2019}).
Generally, $E_{\rm cyc}$ remains stable or shows a positive correlation with luminosity when the luminosity is below a critical value, which can be explained as variations in the emission region height~\citep{Rothschild2017} and/or the Doppler effect \citep{Mushtukov2015a}.
On the other hand, the negative $E_{\rm cyc}$--luminosity correlation is only firmly identified in bright outbursts of V0332+53 and 1A 0535+262, \endnote{but  a possible negative correlation is also seen in the faint states of 1A 0535+262 and GRO J1008-57 \citep{Shui2024, Chen2021}.} \citep{Doroshenko2017, Vybornov2018, Kong2021, Shui2024} (as an example, see Figure~\ref{fig:kong2021}).
A lower $E_{\rm cyc}$ at higher luminosities may arise from the radiation emitted by a higher accretion column or from the reflection process occurring away from the magnetic poles, where the local magnetic field is weaker \citep{Poutanen2013}.
In addition, in the supercritical regime, both the Doppler effect due to the bulk motion and the gravitational redshift should be considered \citep{Loudas2024b}.
We note that the transition between positive and negative $E_{\rm cyc}$ correlations serves as a key indicator for confirming the onset of the accretion column (for details, see Section~\ref{poles}).

\textls[-20]{The long-term evolution of CRSFs has been reported in Her X-1 and \mbox{Vela X-1  \citep{Staubert2014,Staubert2016,Staubert2020, Xiao2019,Xiao2024,Klochkov2015,LaParola2016,Ji2019}}}.
Both sources exhibit an $E_{\rm cyc}$ decay period ($\dot{E}_{\rm cyc}$ $\sim$ several 0.1\,keV per year) and an $E_{\rm cyc}$-invariant period over a timescale of several decades.
The long-term behavior likely reflects a geometric displacement of the line-forming region and/or distortions of the magnetic field around polar caps.
Considering the rapid decay rate and the sudden jump of $E_{\rm cyc}$ between 1990 and 1995 in Her X-1, the long-term evolution of $E_{\rm cyc}$ probably follows a cyclic pattern, which should be confirmed in the future \citep{Staubert2014}.

\section{Polarimetry} \label{polarimetry}
The X-ray emissions of AXRPs were initially presumed to be strongly polarized due to the effects of the plasma and vacuum birefringence  as the radiation propagates through a strongly magnetized plasma. These properties made AXRPs ideal candidates for polarization measurements \citep{Nagel1981, Meszaros1988, Caiazzo2021, Caiazzo2021b}. The polarization degree (PD) and polarization angle (PA) are highly sensitive to the geometry of the emission regions and magnetic fields, offering a unique opportunity to study radiative transport and matter dynamics in the immediate vicinity of pulsars. The launch of the first highly sensitive X-ray polarimeter \ixpe marked a significant step in this field, allowing researchers to test these theoretical predictions. Over its first two years of operation, \ixpe observed 10 AXRPs, and  impressive achievements have been presented in the latest review \citep{Poutanen2024}. Here, we just repeat some of key findings.

For all observed AXRPs, the measured PD was very low $\sim$$5\text{--}20\%$, which is significantly lower than the prediction  ($\sim$$60\text{--}80$ \%) by all existing theoretical models \citep{Doroshenko2022, Tsygankov2023, Zhao2024}. Both PD and PA evolve along with the pulse phase. The PD has a rather complex variation and is anti-correlated with flux. On the other hand, the PA usually has a smooth profile and can be described with the rotating vector model (RVM) (but see \citet{Doroshenko2023} for situations where an unpulsed polarized component exists).  Due to the relatively narrow energy band coverage of \ixpe ($\sim$$2\text{--}8$ keV), a positive relationship between the PD and the photon energy was observed in only a few sources, such as X~Persei \citep{Mushtukov2023} and Vela X-1 \citep{Forsblom2023}. For most other AXRPs, however, both PD and PA appear largely energy-independent. In addition, a $\sim$90$^{\circ}$ difference in the PA has been found between 2 and 3\,keV and above 5\,keV in Vela X-1, which may be caused by vacuum resonance \citep{Forsblom2023,Lai2003}.
As the era of X-ray polarimetry begins, further detailed polarization measurements at varying states, enabled by \ixpe and the forthcoming {\it eXTP} observatory \citep{Zhang2016}, promise to deliver transformative insights. 

\section{Related Binary Systems}\label{Related}

Observations across multiple wavelengths have led to the discovery of intriguing types of binary systems beyond AXRPs.  Here, we only provide very short discussions on two of the most striking binary systems---gamma-ray binaries and PULXs. 

A small yet growing number of HMXBs are found to display bright emissions in nearly all wavelengths (from radio to TeV), being distinguished from the majority of HMXBs. The radiation from these sources with orbital modulation is relatively stable except when the compact objects pass through the periastron. Due to their high-energy output, predominantly above 1 MeV, they are classified as ``gamma-ray binaries'' \citep{Dubus2013}. The mechanisms driving particle acceleration and high-energy emissions in these systems remain poorly understood. Key breakthroughs have been achieved with the detection of radio pulsations from PSR~B1259-63 (48 ms), PSR~J2032+4127 (143 ms), and LS~I~+61$^{\circ}$~303 \mbox{(269 ms) \citep{Johnston1992, Camilo2009, Weng2022}}, and  gamma-ray pulsations have also been reported for PSR~J2032+4127 \citep{Camilo2009}. These results clearly confirm the fact that broadband emissions come from shocks formed by the collision of pulsar and stellar winds, ruling out the accretion-powered microquasar scenario. The most essential difference between the AXRPs reviewed in this paper and gamma-ray binaries is that the former is accretion-powered, while the latter is mostly a non-accreting system, although the nature of the other gamma-ray binaries is still under debate.

To date, more than about 1800 ULXs with $L_{\rm X} \sim 10^{39}\text{--}10^{42}$ erg~s$^{-1}$ have been discovered in nearby galaxies \citep{Walton2022}. The most relevant result was the discovery of the X-ray pulsations from M82~X-2 in 2014, which identified it as the first PULX \citep{Bachetti2014}. Since then, X-ray pulsations have only been detected from about less than 10 PULXs after extensive \mbox{searching \citep{Furst2016,Israel2017a,Israel2017b, Carpano2018, Sathyaprakash2019,Rodriguez2020}}. Nevertheless, the orbital periods and superorbital periods were also reported for several sources, e.g., M82~X-2 \citep{Bachetti2014}, NGC~5907~ULX-1 \citep{Walton2016}, and M51~ULX-7 \citep{Rodriguez2020, Brightman2020}. The periodic signal of $\sim 64$ days detected from NGC~7793~P13 remains contentious, with debates about whether it represents the orbital or superorbital period \citep{Motch2014, Hu2017, Furst2018}.  As can be seen in Figures \ref{fig:corbet} and \ref{fig:super}, PULXs and disk-fed SGXBs located at the same regions in both diagrams, indicating the presence of RLO in PULXs. However, care should be taken as the sample of PULXs is very small.

{PULXs share s similar spin-up trend with the Galactic AXRPs} but with much higher luminosity and spin-up rates \citep{King2023}. Therefore, many works would like to address these two types of sources together, calling both ``PULXs''.  But, actually, the canonical PULXs in outer galaxies are sharply distinct from Galactic AXRPs in both spectral and temporal properties: 1. Galactic AXRPs can only occasionally reach the super-Eddington luminosity ($\sim$$10^{39}$~erg~s$^{-1}$), while PULXs can stay at the ultraluminous state for a long time with much higher luminosity. 2. In contrast to the complex pulse profiles emerged in AXRPs (e.g., Figure \ref{fig:a0535_profile}), the pulse profiles of all PULXs are nearly sinusoidal. 3. The high signal-to-noise spectra of PULXs display a curvature of around 10 keV and can be phenomenologically described by a two-component (or three-component) model. Apart from the details of diverse models, the radiation of PULXs is usually dominated by soft X-rays \citep{Kaaret2017,Koliopanos2017} and tends to become harder when brighter \citep{Weng2018}. However, the spectra of AXRPs are significantly harder with a photon index of $\Gamma < 1.0$ and move along the diagonal branch \citep{Reig2013}. These observational results suggest that the X-ray radiation in PULXs is not strongly beamed, and the accretion flows have different geometries in PULXs and AXRPs \citep{Mushtukov2015c, Mushtukov2021, King2023}. As we discussed for AXRPs, the strength of magnetic fields and the type of optical donors are two key factors in NS accreting systems. These significant differences in observational phenomena are thought to be due to the differences in the strength of their magnetic fields and/or the accretion rate in PULXs and AXRPs. There is still a huge controversy regarding whether PULXs have magnetar-like ($B \geq 10^{13}$ G) or normal magnetic fields as AXRPs ($B \sim 10^{12}$ G) \citep{Eksi2015, Tong2015, Tsygankov2016, King2019, Erkut2020}, though potential CRSFs were reported \citep{Brightman2018, Walton2018,Middleton2019}. Additionally, the donor stars in PULXs are intriguing but challenging to study due to their distance. For example, the companion star of NGC~7793~P13 was identified as a B9Ia supergiant \citep{Motch2011}, and a red supergiant donor star was suggested for NGC~300~ULX-1 \citep{Heida2019}. Further research is necessary to elucidate these enigmatic systems.

\section{Multi-Wavelength Advances}\label{Multi}
Although the primary radiation of AXRPs is emitted in the X-ray band, a multi-frequency approach is clearly essential to fully understand the physical mechanism driving their behaviors.
For instance, in the era of \gaia, accurate measurements of the position, distance, and velocity of HMXBs can be obtained. 
This allows us to estimate their luminosities precisely and enables constraints on natal kicks, birthplaces, and the spatial distribution of NSs \citep{Bailer2021, Fortin2022, Fortin2022b}.
On the other hand, optical observations have been performed for investigating the properties of companion stars and directly reflecting the inner-binary environment, such as the truncated and warped decretion disks \citep{Reig2018, Reig2022}.

Recent radio observations strikingly detect the radio counterparts of AXRPs, and the $L_{\rm x}$-$L_{\rm radio}$ plane shows a correlation as $L_{\rm radio} \propto L_{\rm x}^{0.86\pm0.06}$ during giant outbursts \citep{Eijnden2018a,Eijnden2018b,Eijnden2019a,Eijnden2022,Eijnden2024a,Eijnden2024b}.
This power-law index is similar to those measured in NS and BH LMXBs, but the radio intensity is significantly lower by 2--4 orders of magnitude. 
Current observations suggest that this radio radiation is likely attributed to jets.
In theory, however, accretion systems with high magnetic fields, such as AXRPs, are not expected to launch jets, and the underlying emission mechanism remains unknown.

Theoretically, AXRPs might also emit MeV radiation.
This is because, in some cases, the kinetic energy of infalling matter exceeds the binding energy of nucleons within nuclei.
This leads to the nuclear spallation of a large portion of the accreted helium and liberates neutrons.
Eventually, a redshifted line is expected due to proton--neutron recombinations $p + n \rightarrow D + \gamma$ (2.223\,MeV) \citep{Reina1974,Brecher1980,Bildsten1993,Jean2001,Ducci2024}.
As of now, several attempts have been made using \textit{COMPTEL }, \textit{RHESSI}, and \textit{INTEGRAL}/SPI, which however only result in upper limits \citep{McConnell1997,Boggs2006,Teegarden2006,CaliSkan2009,Ducci2024}.
The detection of MeV lines from accreting pulsars remains one of the key scientific projects for future MeV astronomy projects, such as \textit{COSI}, \textit{eASTROGAM}, and \textit{MASS} \citep{Tomsick2022,Angelis2018,Zhu2024}.

%
%
\vspace{6pt} 





\authorcontributions{
Sections~\ref{intro},~\ref{temporal},~\ref{polarimetry} and~\ref{Related} were primarily contributed by Shan-Shan Weng and Sections~\ref{model},~\ref{spec} and~\ref{Multi} were mainly contributed by Long Ji.
All authors have read and agreed to the published version of the manuscript.
}

\funding{
This work is supported by National Natural Science Foundation of China under grants No. 12473041, 12173103, and 12261141691.
}

\dataavailability{
All data used in this paper are public from other cited papers or the High Energy Astrophysics Science Archive Research Center (HEASARC) \url{https://heasarc.gsfc.nasa.gov}.
}

\acknowledgments{We thank the anonymous referees for their comments and suggestions, which improved the quality of this paper. The authors thank the National Natural Science Foundation of China for their support under grants 12473041, 12173103, and 12261141691. }

\conflictsofinterest{The authors declare no conflicts of interest.} 



\abbreviations{Abbreviations}{
The following abbreviations are used in this manuscript:\\

\noindent 
\begin{tabular}{@{}ll}
{AXRPs}                                 & Accreting X-ray pulsars   \\
{BeXBs}                               & Be X-ray binaries \\
{BHs}				   & Black holes \\
{BFM}				   & Beat-frequency model \\
{CRSFs}                               & Cyclotron resonant scattering features \\
{GPD}                                      & Gas-pressure-dominated\\
{HID}    &                                Hardness-intensity diagram \\
{HMXBs}                              & High-mass X-ray binaries   \\
{IMXBs}                              & Intermediate-mass X-ray binaries   \\
{KFM}                                   & Keplerian-frequency model    \\
{LMC}                                   & Large Magellanic Cloud  \\
{LMXBs}                                   & Low-mass X-ray binaries  \\
{NSs}				   & Neutron stars \\
{PA}                                     & Polarization angle  \\
{PD}                                     & Polarization degree  \\
{PULXs}				 & Pulsating ultraluminous X-ray sources \\
{QPOs}				   & Quasi-periodic oscillations \\
{RLO}				& Roche-lobe overflow \\
{RPD}                                      &  Radiation-pressure-dominated\\
{RVM}				  &  Rotating vector model \\
{SFXTs}                               & Supergiant fast X-ray transients  \\
{SGXBs}                              & Supergiant X-ray binaries  \\
{SMC}                                  & Small Magellanic cloud  \\
{ULXs}                                 & Ultraluminous X-ray sources  \\
{XRBs}                                 & X-ray binaries  \\
\end{tabular}
}

\appendixtitles{yes} 
\appendixstart
\appendix
%
%

\begin{adjustwidth}{-\extralength}{0cm}
\printendnotes[custom] 

\reftitle{References}

\PublishersNote{}
\end{adjustwidth}
\end{document}